\documentclass[twocolumn, pra, showpacs, superscriptaddress, floatfix, longbibliography]{revtex4-1}

\usepackage[utf8]{inputenc}
\usepackage{amsmath}
\usepackage{amssymb}
\usepackage{graphicx}
\usepackage{float}
\usepackage{hyperref}
\usepackage{xcolor}
\usepackage{placeins} 
\usepackage{booktabs} 

\hypersetup{
    colorlinks=true,
    linkcolor=blue,
    citecolor=blue,
    urlcolor=blue,
}

\begin{document}

\title{Modeling Protein Evolution via Generative Inference \\ 
From Monte Carlo Chains to Population Genetics}

\author{Leonardo Di Bari}
\affiliation{DISAT, Politecnico di Torino, Corso Duca degli Abruzzi, 24, I-10129, Torino, Italy}
\affiliation{Sorbonne Universit\'e, CNRS, Laboratory of Computational and Quantitative Biology, 75005 Paris, France}

\author{Thierry Mora}
\affiliation{Laboratoire de Physique, \'Ecole Normale Sup\'erieure, CNRS, PSL Universit\'e, Sorbonne Universit\'e, Universit\'e de Paris, 75005 Paris, France}
\affiliation{The James Franck Institute, The University of Chicago, Chicago, Illinois 60637, USA}
\affiliation{Department of Physics, The University of Chicago, Chicago, Illinois 60637, USA}

\author{Andrea Pagnani}
\affiliation{DISAT, Politecnico di Torino, Corso Duca degli Abruzzi, 24, I-10129, Torino, Italy}
\affiliation{INFN, Sezione di Torino, Via Pietro Giuria 1, Torino 10125, Italy}
\affiliation{Italian Institute for Genomic Medicine, IRCCS Candiolo, SP-142, Candiolo 10060, Italy}

\author{Aleksandra M. Walczak}
\affiliation{Laboratoire de Physique, \'Ecole Normale Sup\'erieure, CNRS, PSL Universit\'e, Sorbonne Universit\'e, Universit\'e de Paris, 75005 Paris, France}
\affiliation{The James Franck Institute, The University of Chicago, Chicago, Illinois 60637, USA}
\affiliation{Department of Physics, The University of Chicago, Chicago, Illinois 60637, USA}

\author{Francesco Zamponi}
\thanks{Corresponding authors. Email: francesco.zamponi@uniroma1.it, saverio.rossi@uniroma1.it}
\affiliation{Dipartimento di Fisica, Sapienza Universit\`a di Roma, Piazzale Aldo Moro 5, 00185 Rome, Italy}

\author{Saverio Rossi}
\thanks{Corresponding authors. Email: francesco.zamponi@uniroma1.it, saverio.rossi@uniroma1.it}
\affiliation{Dipartimento di Fisica, Sapienza Universit\`a di Roma, Piazzale Aldo Moro 5, 00185 Rome, Italy}

\date{\today}

\begin{abstract}
Generative models derived from large protein sequence alignments define complex fitness landscapes, but their utility for accurately modeling non-equilibrium evolutionary dynamics remains unclear. In this work, we perform a rigorous comparative analysis of three simulation schemes, designed to mimic evolution {\it in silico} by local sampling of the probability distribution defined by a generative model.
We compare standard independent Markov Chain Monte Carlo, Monte Carlo on a phylogenetic tree, and a population genetics dynamics, benchmarking their outputs against deep sequencing data from four distinct {\it in vitro} evolution experiments. We find that standard Monte Carlo fails to reproduce the correct phylogenetic structure and generates unrealistic, gradual mutational sweeps. Performing Monte Carlo on a tree inferred from data improves phylogenetic fidelity and historical accuracy. 
The population genetics scheme successfully captures phylogenetic correlations, mutational abundances, and selective sweeps as emergent properties, without the need to infer additional information from data.
However, the latter choice come at the price of not sampling the proper generative model distribution at long times. 
Our findings highlight the crucial role of phylogenetic correlations and finite-population effects in shaping evolutionary trajectories on fitness landscapes. These models therefore provide powerful tools for predicting complex adaptive paths and for reliably extrapolating evolutionary dynamics beyond current experimental limitations.
\end{abstract}

\maketitle

\section{Introduction}

Proteins are fundamental building blocks of life, performing a vast array of essential biological functions. Over evolutionary timescales, mutations in the genetic code can alter the amino acid sequence of a protein, potentially affecting its stability, structure, or function. Accurately modeling protein evolution is therefore of both theoretical and practical importance. From a theoretical perspective, realistic models are crucial for phylogenetic inference and for understanding the constraints that shape and structure evolutionary landscapes. From a practical standpoint, such models are essential for predicting the evolutionary trajectories of pathogens and for informing the design of effective drug therapies.

The modeling of molecular evolution fundamentally relies on statistical frameworks to describe how protein or nucleic acid sequences change over time. 
Historically, this effort has been motivated by the need to understand the forces of natural selection and to infer evolutionary relationships, leading to the development of sequence substitution models~\cite{pupko2020agentle}.
These models provide the core mathematical framework for comparative genomics and phylogenetic inference. 
They treat the evolution of a sequence as a continuous-time Markov process, where a transition rate matrix governs the rate of change from one nucleotide (or amino acid) to another. 
The complexity of these models ranges widely: the simplest, like the Jukes–Cantor (JC69) model for nucleotides~\cite{jukes1969evolution}, assumes that all substitutions occur at the same rate. 
More sophisticated models, such as HKY85 and GTR for DNA~\cite{Kimura1980asimple,Tavar1986Some}, and WAG~\cite{whelan2001ageneral}, JTT~\cite{Jones1992TheRG} and LG~\cite{Le2008animproved} for proteins, account for biases like transition/transversion rates and unequal stationary frequencies. 

While these substitution models have been very successful, they mostly operate under a severe simplification: the mutation rate of each site does not depend on the rest of the sequence. 
This is equivalent to assuming that sites evolve independently.
In order to reproduce statistical properties of evolving sequences, such as the distribution of substitution rates among sites or the time-dependent mutability constraints at different residues, one then needs to add empirical observations explicitly. 
More generally, these standard models are inherently unable to capture epistatic interactions---where the functional effect of one mutation depends on the context in which it occurs---that are crucial for maintaining protein structure and function, and strongly influence evolutionary trajectories~\cite{weinreich2005perspective, lunzer2010pervasive, starr2016epistasis, otwinowski2018biophysical, otwinowski2018biophysical,  domingo2019causes, phillips2021binding, bakerlee2022idiosyncratic,
buda2023pervasive,johnson2023epistasis}.
Considering site interactions makes the algorithm more computationally expensive and prevents the use of transition matrices, but it also makes it more accurate. 
One attempt at including epistasis into evolutionary modeling was done with structural substitution models~\cite{arenas2015,bordner2014}, in which an interaction between sites mediated by the structure is taken into consideration.

Here, we consider an alternative approach in which first a fitness landscape is inferred from natural sequence data using generative sequence modeling, and then the evolutionary dynamics is simulated on that landscape. 
The advantage of working directly in sequence space, without explicitly incorporating structural information, is that the resulting models are computationally very simple and efficient.
We employ a powerful energy-based generative modeling framework known as Direct Coupling Analysis (DCA)~\cite{cocco2018inverse}. 
DCA is trained on a dataset of natural sequences, from which it infers a probability distribution with parameters representing site-specific constraints and, critically, the long-range epistatic interactions that stabilize protein structure and dictate function~\cite{ferguson2013translating,figliuzzi2016coevolutionary,levy2017potts,couce2017mutator,vigue2023predicting}. 
DCA-based models have already proven useful in various tasks, including contact prediction~\cite{Morcos2011,marks2011protein}, the \textit{in silico} generation of  functional proteins~\cite{russ2020evolution, Calvanese2025generating, Lambert2024expanding}, the study of specificity-switching pathways~\cite{Rehan2025design} and the inference of mutational fitness landscapes from experiments~\cite{figliuzzi2016coevolutionary,cossio2021unsupervised,Sesta2021amala, chen2023understanding,Sesta2024inference,DeLeonardis2024unsupervised}. Similar approaches have also been used in the context of HIV/HCV~\cite{barton2016relative, flynn2017inference, Hart_2019, choudhuri2022contingency, biswas2024kinetic} and SARS-CoV-2~\cite{Huot2025constrained, Huot2025generative}. Hence, sequence fitness landscapes inferred by DCA contain information about structure, and are able to capture the main ingredients needed to make a sequence functional, as well as the epistatic effect of mutations.
While alternative generative modeling approaches to fitness landscapes exits, e.g. based on Large Language Models, to our knowledge DCA is not 
outperformed by them on the modeling tasks listed above, hence we stick
to it due to its simplicity of implementation and computational 
efficiency~\cite{Rosset2026,bereux2024fast,fernandez2024accelerated}.

While DCA has traditionally been applied at equilibrium---generating novel protein sequences by sampling the inferred probability distribution via independent Markov chains, for times much larger than the mixing time of the model---its use has increasingly expanded to forward evolutionary simulations, where its energy is interpreted as a fitness function~\cite{Biswas2019,de2020epistatic,bisardi2022modeling,alvarez2024vivo,alvarez2022novel,biswas2024kinetic,DiBari2024,rossi2024fluctuations}. 
Previous implementations are limited in realism because they assume independence among individual trajectories, thereby neglecting phylogenetic correlations and population size effects.
In addition, these models often work in discrete time, limiting the possibility of including them in standard pipelines for phylogenetic inference.
Nevertheless, it has been recently shown that it is possible to inform and refine ancestral sequence reconstruction~\cite{DeLeonardis2024reconstruction} through a DCA-inferred landscape, and that the sampling of such landscape can also be performed through a phylogenetically-suited continuous time model~\cite{Pagnani2025generative}. This validates the use of DCA-inferred fitness landscapes as the basis for modern, interpretable and epistasis-aware evolutionary models.

At the same time, the experimental study of protein evolution has seen substantial advances. 
Over the past two decades, high-throughput experimental techniques have emerged, dramatically enhancing our ability to directly probe functional fitness landscapes.
Techniques like deep mutational scanning (DMS)~\cite{fowler2014deep,  romero2015dissecting, GONZALEZ2019fitness} systematically measure the mutational effect of a vast number of mutations across a protein domain, one by one.
The effect of multiple mutations is also under study in combinatorial landscapes, in which all possible variants between two proteins are tested~\cite{buda2023pervasive,Poelwijk2019,westmann2024entangled}. 
These quantitative, high-resolution functional maps reveal that epistatic interactions are not just present, but are a pervasive feature of molecular fitness.
{\it In vitro} evolution~\cite{DENG2012deep, melnikov2014comprehensive,kitzman2015massively,fantini2020protein, stiffler2020protein,dacosta2023inferring,rix2023continuous} experiments, performed through rounds of mutagenesis, selection and amplification, provide a glance at the trajectories resulting from the exploration of fitness landscapes starting from a specific wild-type.  
The availability of this rich, detailed experimental data now makes it possible to rigorously assess the ability of evolutionary models to reproduce real-world biological results, especially those involving complex epistasis.

Building upon these advances in both experimental techniques and modeling approaches, in this work we seek to rigorously evaluate the performance and inherent limitations of DCA-informed simulation approaches used to model high-throughput evolution experiments. 
Specifically, we investigate and compare three distinct computational schemes to quantitatively reproduce experimental data. 
This comparative study addresses a critical theoretical gap: understanding the performance differences between simple, independent-chains Markov Chain Monte Carlo (MCMC) approaches and more detailed, computationally intensive population genetics models. 
By identifying the specific conditions under which the minimal Monte Carlo framework proves sufficient, and what population-level dynamics it neglects, we aim to provide clarity for future computational studies. 
Furthermore, developing highly predictive and reliable \textit{in silico} simulation tools is of significant practical value for prioritizing experimental efforts and advancing protein engineering.
The study is structured as follows: we first describe the integration of DCA-derived fitness landscapes with four time-resolved experimental datasets.
We then introduce three simulation frameworks operating on the inferred fitness landscapes, presented in order of increasing complexity: standard MCMC, phylogenetically constrained MCMC, and a discrete-generation population genetics model. We evaluate how well the models reproduce empirical sequence diversity and compare them to observed phylogenetic structure and site frequency spectra to assess historical fidelity. 
Finally, we show the individual mutational trajectories to determine which framework best captures the non-equilibrium dynamics of selective sweeps, concluding with a discussion on the long-term predictive power of these models beyond current experimental limitations.

\begin{figure*}
    \centering
    \includegraphics[width=0.8\textwidth]{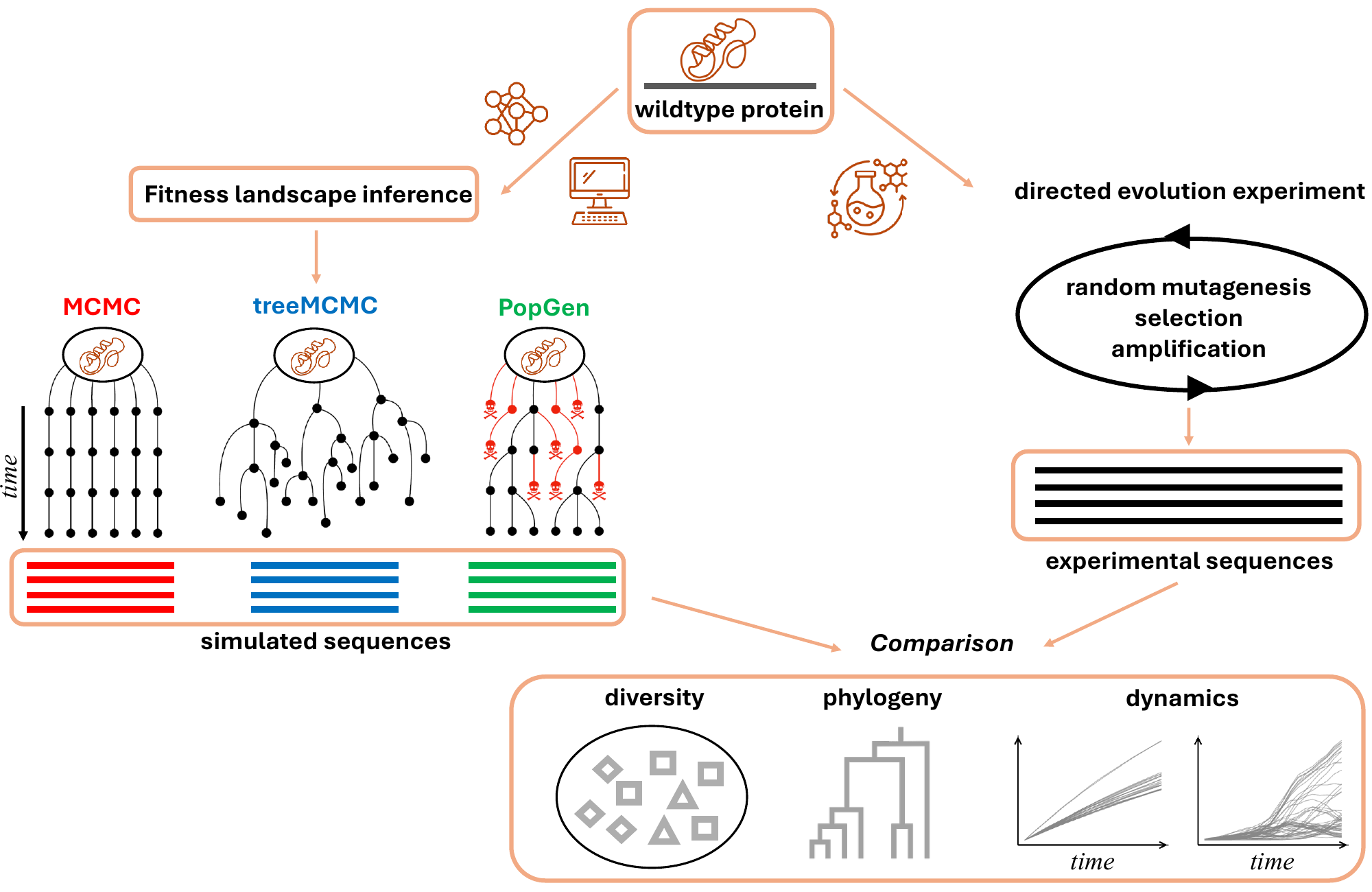}
    \caption{\textbf{Schematic of the experimental and simulation pipeline.} 
{\it In vitro} evolution data are used to benchmark three evolutionary simulation schemes (MCMC, treeMCMC, PopGen) over four different evolution experiments realized through rounds of mutagenesis, selection and amplification. 
Simulated and experimental sequences are compared through divergence, diversity, phylogeny, and mutational spectra.}
\label{fig::Scheme}
\end{figure*}

\section{Materials and Methods}

The graphical abstract in Fig.~\ref{fig::Scheme} outlines the workflow used to compare synthetic evolutionary schemes with experimental evolution ones. 
Evolution experiments begin from a wild-type protein and proceed through iterative cycles of mutagenesis, selection, and amplification, generating time-resolved sequence populations measured by deep sequencing.
These experiments can be carried on in different evolutionary regimes.
Here we focus on experiments that performed evolution under a moderate selection strength, such that the initial wild type would survive. 
Hence, they performed neutral space exploration, instead of directed evolution that aims at maximizing fitness under very strong selection pressure.
For the training of our generative model, we start by gathering natural homologs of the wild-type protein into a curated Multiple Sequence Alignment (MSA).
This MSA is then used to infer a pairwise Potts model, providing an energy-based description of the fitness landscape, through standard DCA techniques.
We exploit the inferred landscape by running three simulation schemes: independent MCMC, MCMC along a phylogenetic tree, and a population-genetics scheme incorporating mutation, selection, and finite-population sampling. 
Each framework produces synthetic evolutionary trajectories. 
Simulated and experimental populations are then compared through their sequence diversity, inferred phylogenetic structure, and DCA energy statistics, allowing us to assess how closely each model captures the dynamics observed in experimental evolution.

\subsection{Experimental Datasets}
Our comparative analysis draws on four time-resolved deep-sequencing datasets generated through multi-round {\it in vitro} evolution experiments. 
These protocols mimic natural selection by iterating cycles of mutagenesis, functional selection, and amplification, thereby enabling the monitoring of how the statistical properties of sequences change across rounds. 
In each experiment, a diversified mutant library derived from a wild-type gene seeds the first round. 
Variants are then subjected to mutations through error-prone PCR, and selection through growth of bacterial cultures containing the plasmid libraries under fixed concentration of antibiotics.
Finally, surviving sequences are amplified to initiate the next cycle. Deep sequencing of populations provides a quantitative picture of the evolutionary process at different experimental rounds.
The first two analyzed datasets~\cite{fantini2020protein, stiffler2020protein} correspond to the $\beta$-lactamase family and include the well-characterized TEM1 and PSE1 enzymes, both evolved under antibiotic selection (ampicillin in this case). 
For TEM1, sequences containing more than six gaps were discarded during quality control, resulting in a curated MSA of sequences sampled from the last generation (round $12$) of the experiment. 
The PSE1 dataset was curated using the same filtering procedure and collected after round $20$. 
A third dataset derives from the AAC6~\cite{stiffler2020protein} acetyltransferase, evolved under functional selection (kanamycin) linked to its enzymatic activity and sequenced at final round $8$.
Finally, we include a directed-evolution experiment performed on dihydrofolate reductase (DHFR)~\cite{dacosta2023inferring}, in which repeated mutagenesis and selection rounds (under trimethoprim) yielded large libraries of functional variants sequenced at round $15$.
Intermediate rounds of experiments where also sequenced in some cases (round $10$ for PSE1, round $6$ for TEM1 and rounds $\left(1,2,3,4,5\right)$ for DHFR). In this work we will focus mainly on studying the final rounds of such experiments, as they show a larger sequence variability, but the same approach can be carried out with each sequenced round of choice. Aligned amino-acid sequences and raw reads are publicly available from the authors’ repository and were curated following the same procedure applied to the other families. 

\subsection{DCA-derived fitness landscape}

For all protein families, natural MSAs were constructed by running \texttt{hmmsearch} from the HMMer suite against UniProt. 
Insertions were removed, sequences with more than 20\% of gapped positions and duplicates were filtered out, as well as those exceeding a predefined $80\%$ similarity threshold to the corresponding wild-type.
Alignments were then restricted to Pfam-annotated positions: the PF13354 $\beta$-lactamase2 family for the wild types TEM1 and PSE1, the PF00583 Acetyltransf1 family for the wild type AAC6, and the PF00186 Dihydrofolate reductase family for the wild type mDHFR.
Wild-type and evolved sequences were aligned to the MSA using \texttt{hmmalign}, and columns corresponding to unmatched states in the wild-type alignment were removed from all experimental MSAs.

All simulations reported in this work take place on a fitness landscape defined by the statistical energy $E(\boldsymbol{a})$ of a pairwise Potts model inferred from the MSA of the natural sequences, where $\boldsymbol{a}$ represents the evolving sequence. 
Following the standard DCA formulation, sequences are assigned a probability
\begin{equation}
\label{eq:prob}
    P(\boldsymbol{a})=\frac{1}{Z}\exp\big[- \beta E(\boldsymbol{a})\big],
\end{equation}
with
\[ E(\boldsymbol{a})=-\sum_{i} h_i(a_i) - \sum_{i<j} J_{ij}(a_i,a_j),
\]
where the site-specific fields $h_i$ and pairwise couplings $J_{ij}$ encode sites conservation and coevolution, respectively. 
The parameter $\beta = 1/T$ is an inverse effective temperature, which can be considered as a proxy for selection pressure; it is generally set as $\beta=1$ during training, and it is tuned during the simulation to match experimental data~\cite{bisardi2022modeling,DiBari2024}.
The statistical energy $E$ is here interpreted as a \emph{negative} proxy for  fitness: sequences with low $E$ are assigned a higher probability and thus predicted to be more functional, whereas high-energy sequences populate fitness valleys and have low functionality.  
Parameters are inferred from the natural sequences via Boltzmann-machine learning (bmDCA) in order to maximize the model likelihood under one- and two-site empirical constraints; this inference procedure yields Potts models that are generative and able to reproduce very well the statistical properties of the training MSA~\cite{figliuzzi2018pairwise}. 
For the present study we employ the production-grade implementation \texttt{adabmDCA\;2.0}~\cite{Rosset2026}, which builds on earlier adaptive Boltzmann-machine strategies~\cite{muntoni2021adabmdca,figliuzzi2016coevolutionary} and provides efficient and parallelized training routines suitable for large MSAs and modern HPC/GPU architectures. 

\subsection{Codon accessibility}

\begin{figure*}
    \centering
    \includegraphics[width=\textwidth]{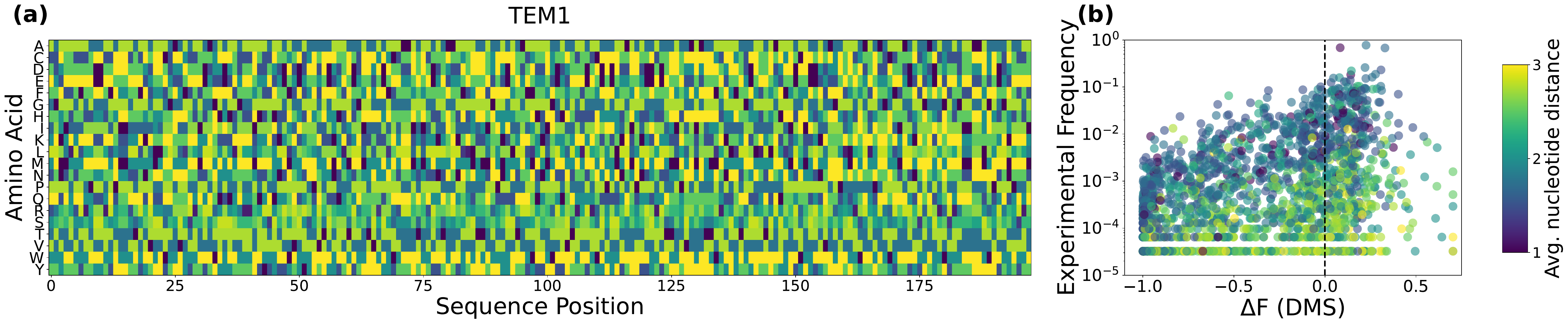}
    \caption{\textbf{Comprehensive mutational and fitness landscape of TEM1.} 
    (a) Nucleotide accessibility map for TEM1. The heatmap shows the average nucleotide distance from each wild-type codon to all possible codons corresponding to a given amino acid substitution. 
    (b) Correlation between experimental variant frequencies in the evolution experiment of Ref.~\cite{fantini2020protein} (round 12) and deep mutational scanning (DMS) fitness measurements ($\Delta F$). Each data point corresponds to a single amino acid mutation. Data points are colored according to their average nucleotide distance from the wild-type sequence. The $\Delta F$ values were obtained from DMS studies of TEM1 variants~\cite{GONZALEZ2019fitness}, with the vertical dashed line at $\Delta F = 0$ demarcating the boundary between beneficial and deleterious mutations.}
    \label{fig::accessibility}
\end{figure*}

DCA modeling works on amino acid sequences, as this is what is often provided by public databases.
However, mutations happen at the level of nucleotides, hence the constraints of protein evolution are also determined by the underlying architecture of the genetic code. 
To capture the earliest stages of divergence from the wild-type, it is thus essential to model mutations at the nucleotide level rather than treating all amino-acid changes as equally accessible. 
While this choice has no consequences at long time-scales, the codon accessibility map imposed by the nucleotide sequence of the wild-type strongly influences the emerging mutations in experiments~\cite{gunnarsson2023predicting}, where the evolving sequences remain quite close to the wild-type. 

We illustrate this constraint for TEM1 in Fig.~\ref{fig::accessibility}a, which displays the average nucleotide distance required to go from the wild-type codon to each possible amino-acid state. 
This map shows that certain residues are significantly less reachable, requiring multiple nucleotide substitutions within a single codon to be expressed. 
This architectural constraint has profound effects on evolutionary trajectories; as previously demonstrated~\cite{gunnarsson2023predicting}, amino-acid accessibility can overcome the fitness advantage of individual mutants at short time scales, leading to fitter variants being less represented than less fit ones simply due to lower initial accessibility.
We observe this exact pattern in our analysis of TEM1 in Fig.~\ref{fig::accessibility}b. 
By correlating experimental single-mutant frequencies against the fitness effects ($\Delta F$) obtained through deep mutational scanning~\cite{GONZALEZ2019fitness}, it becomes evident that variant representation is heavily biased by the mutational distance from the wild-type. 
Variants requiring nearly a single nucleotide change (shown in dark blue/purple) appear more frequently in the evolution experiment compared to those at bigger distance (green/yellow), even when the latter possess comparable or superior fitness effects. 
These findings strongly suggest that one should simulate evolution experiments by performing mutations in nucleotide space and selection on the available fitness landscape in amino-acid space~\cite{bisardi2022modeling,DiBari2024}, a choice we follow in the rest of this work.

\subsection{Simulation schemes}

Our first simulation scheme, MCMC, models protein evolution as a stochastic process. We simulate sequence dynamics at the codon level, explicitly exploiting the structure of the genetic code. 
This approach has already been shown to reproduce the statistical properties observed in evolution experiments~\cite{bisardi2022modeling,DiBari2024}. 
Evolution is defined on nucleotide sequences of $L$ codons, $\mathbf{c} = (c_1,\dots,c_L)$, which are mapped to amino acid sequences $\mathbf{a}(\mathbf{c})$ via the genetic code, allowing us to naturally account for point mutations and insertion/deletion events.
The evolutionary dynamics is guided by an energy function defined at the nucleotide level,
\begin{equation}
E(\mathbf{c}) = E[\mathbf{a}(\mathbf{c})] - T \sum_{i=1}^{L} \log f(c_i \mid a_i) \ ,
\end{equation}
which extends the DCA-derived amino acid energy by incorporating codon usage bias.
Here, $f(c_i \mid a_i)$ is the occurrence frequency of codon $c_i$, coding for amino acid $a_i$ on site $i$. 
Here, for simplicity, we assume no codon bias and we assign equal probability to all the codons that code for amino acid $a_i$, hence $f(c_i \mid a_i) = 1/N_{a_i}$, where $N_{a_i}$ is the number of codons coding for amino acid $a_i$.
Evolutionary trajectories are generated by running independent Markov Chain Monte Carlo (MCMC) simulations that combine single-nucleotide mutations and codon-level insertion/deletion moves. 
Both types of moves are constructed to satisfy detailed balance with respect to the energy $E(\mathbf{c})$, ensuring convergence to the target distribution $P(\mathbf{a})$ of natural protein sequences, as in previous work~\cite{DiBari2024}.

To overcome the absence of phylogenetic correlations in the standard MCMC sampling just defined, we developed a MCMC sampling scheme along an inferred phylogenetic tree, called treeMCMC in the following.
In this approach, we first infer a phylogenetic tree via standard techniques from the experimental data we wish to describe, and then we simulate evolution along the branches.
This imposes a historical constraint on the sequence generation, partially accounting for the time-ordering of substitutions, and introduces correlation between evolving sequences.
To ensure that the simulation scheme correctly reproduces the phylogenetic correlations, the tree is inferred from experimental sequences at the last available round (e.g. round 12 of evolution for TEM1 in Ref.~\cite{fantini2020protein}) exploiting FastTree~\cite{price2010fasttree} within the CAT approximation and the WAG substitution model, which returns a tree together with a set of branch lengths in some arbitrary time units.
Having fixed the tree in this way, our simulated evolutionary process begins at the root with the wild type sequence, and descends to the leaves. 
The sequence $\mathbf{a}$ at a specific node is defined by evolving its parental sequence using the MCMC algorithm.
The number of attempted mutations (MCMC steps) from the parent to the child sequence is given by $\mu \times l_\mathbf{a} \times L$, where $l_\mathbf{a}$ is the original (fixed) length of the branch connecting node $\mathbf{a}$ to its parent, $\mu$ is a tunable mutation rate (expressed in the inverse of the time unit that defines the branch length) and $L$ is the length of the protein sequence. We repeat this procedure until all the leaf nodes of the tree have been sampled. In order to fix $\mu$, we perform a grid-search in parameter space, which will be later described.
Notice that in the treeMCMC scheme, we actively add phylogenetic information about the experimental sequences in the simulation setup, while in the two other cases (MCMC and PopGen) the evolutionary model is agnostic to experimental results.

The PopGen simulation scheme follows the framework of established evolutionary models~\cite{barton2016relative,Hart_2019, shimagaki2025high}, explicitly simulating discrete rounds of experimental evolution through Wright-Fisher-like dynamics. 
Unlike Monte Carlo-based approaches, this framework captures the non-equilibrium nature of laboratory settings by accounting for finite-size population and species competition effects.
Starting from a population of replicates of the ancestral sequence, each evolutionary round iterates through the following steps.
\begin{itemize}
    \item Mutation: Sequences undergo single-nucleotide mutations at non-gapped positions. The mutation probability at each site $\tilde{p}$ is derived from a generalized Jukes-Cantor model with $q=4$ nucleotides:
    $$\tilde{p} = \frac{3}{4} \left( 1 - e^{-\nu} \right),$$
    where $\nu$ represents the mutation rate. 
    Codons can be replaced by gaps (a single codon containing three gaps) and vice versa with probability ${\tilde{p}' = 1-e^{-\nu'}}$. Once a gap is changed into a codon, a random codon is proposed~\cite{DiBari2024}.

    \item Selection: Each sequence $\boldsymbol{a}$ survives selection with a probability $W(\boldsymbol{a})$, determined by its DCA energy $E(\boldsymbol{a})$ following:
    $$W(\boldsymbol{a}) = \frac{1}{1+e^{\tilde{\beta} \left( E(\boldsymbol{a})-\tilde{E} \right)}},$$
    where $\tilde{E}$ defines the selective threshold and $\tilde{\beta}$ denotes the selection strength.

    \item Amplification: The total population size is restored via multinomial sampling of the survived sequences, completing the cycle.
\end{itemize}
It should be noted that, contrarily to the MCMC and treeMCMC schemes, the PopGen dynamics does not satisfy detailed balance with respect to the DCA probability $P(\mathbf{a})$, hence convergence to a sample from this probability is not guaranteed (and does not occur, as we discuss below).

\subsection{Calibration}

It has been previously shown~\cite{bisardi2022modeling, DiBari2024} that to reproduce the mutational diversity of {\it in vitro} evolution experiments it is necessary to tune the model parameters in order to mimic the correct divergence time and selective pressure, which obviously depend on the experimental setting.
The quantities we chose to look at to summarize these two characteristics are: the average Hamming distance (number of amino acid mutations) of the evolving sequences from the wild-type one, $\langle H\rangle$, and the average DCA energy of the evolving sequences $\langle E \rangle$. 
At short evolutionary times, the longer a sequence evolves, the further away it will be from the wild-type; at longer times, when the evolution reaches equilibrium, the average distance from wild-type saturates to a finite value.
At the same time, as the energy plays the role of negative fitness, the average energy provides a measure of the overall selection strength, that can be matched to that of experimental sequences. 

Because during short {\it in vitro} evolution experiments gaps and insertions are not introduced by construction, in our models we set the mutation rates for gaps to zero. Hence, gaps remain fixed as in the wild type
sequence during evolution.
The key remaining parameters for all three schemes were calibrated against experimental data using a grid search to minimize the loss function $ \mathcal{L} = (\langle H \rangle_{\text{sim}} - \langle H \rangle_{\text{exp}})^2 + (\langle E \rangle_{\text{sim}} - \langle E \rangle_{\text{exp}})^2 $, effectively matching the average Hamming distance and DCA energy. In the MCMC and treeMCMC frameworks, this corresponds to tuning the inverse selective pressure $\beta$ and the mutational distance, which is defined by the number of attempted steps $N_{\rm steps}$ in MCMC and by the mutation rate $\mu$ in treeMCMC. For the latter, the underlying phylogenetic tree was inferred directly from the experimental sequences.
In PopGen we need instead to fix the values of three parameters -- the mutation rate $\nu$, the selection threshold $\tilde{E}$ and the selection strength $\tilde{\beta}$ -- as well as the number of rounds.
The latter is chosen to match that of each specific evolution experiment ($20$ for PSE1, $12$ for TEM1, $8$ for AAC6 and $15$ for DHFR). 
The amount of synthetic sequences produced in all simulations is $N=10^5$, with the exception of treeMCMC, in which we obtain the exact same number of each experiments (due to the fact that we infer the tree from experimental sequences).
The grid-search in parameter space provides a set of optimal parameters that will be used to reproduce the experiments, see the Supplementary Materials for details. 
Obviously, the same procedure can be carried on to tune the parameters in order to fit intermediate rounds. 
More interestingly, one can also opt to reproduce the statistics of intermediate rounds just by using the optimal parameters for the last round and running the simulation for a proportionally shorter time. 
In the Supplementary Materials we show that this is sufficient to obtain a relatively good agreement with round $10$ of PSE1.

\section{Results}

The pipeline illustrated in Fig.~\ref{fig::Scheme} evaluates how effectively different computational models based on realistic fitness landscapes replicate the experimental outcomes of protein evolution. 
On the fitness landscape inferred via Direct Coupling Analysis (DCA) we simulate evolutionary trajectories starting from a wild-type sequence and compare these \textit{in-silico} results to actual experimental populations.
Our analysis centers on whether these models capture the fundamental ``signatures'' of evolution. Specifically, we examine the sequence diversity within the populations, the phylogenetic correlations that define the lineage structure, and the temporal mutational dynamics across successive rounds. 
This comparison highlights successes and criticalities of each simulation framework in reproducing the non-equilibrium and historical constraints observed in laboratory settings.

\begin{figure}[t]
    \centering
    \includegraphics[width=0.8\columnwidth]{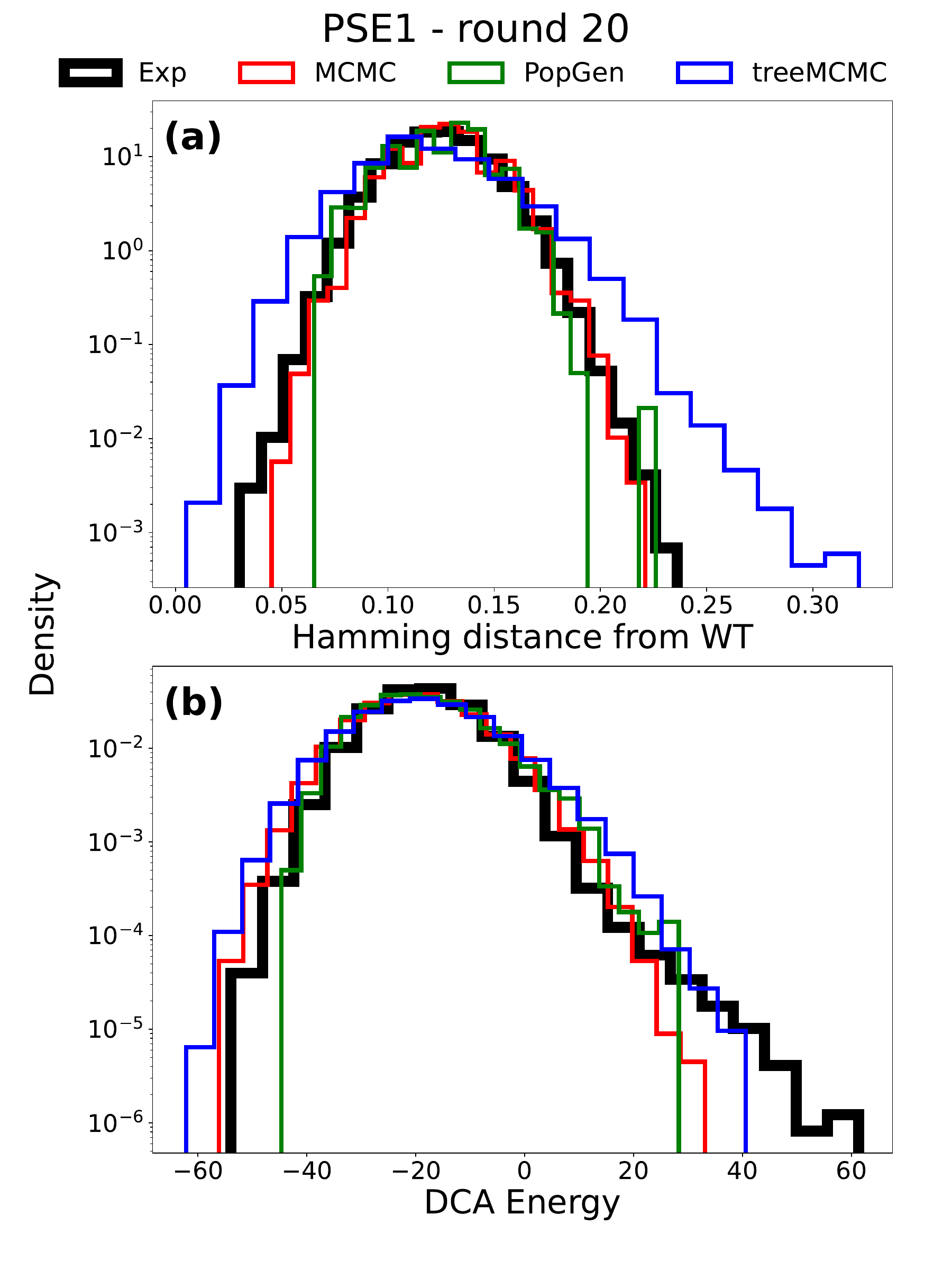}
    \caption{\textbf{Tuning of simulation schemes for the PSE1 experiment.} The distribution of (a) fractional Hamming distance $H/L$ from the wild-type and (b) DCA energy $E$ for the experimental PSE1 population compared to the three simulation schemes after parameter tuning.
    }
    \label{fig::tuning}
\end{figure}

\begin{figure*}
    \centering
    \includegraphics[width=0.8\textwidth]{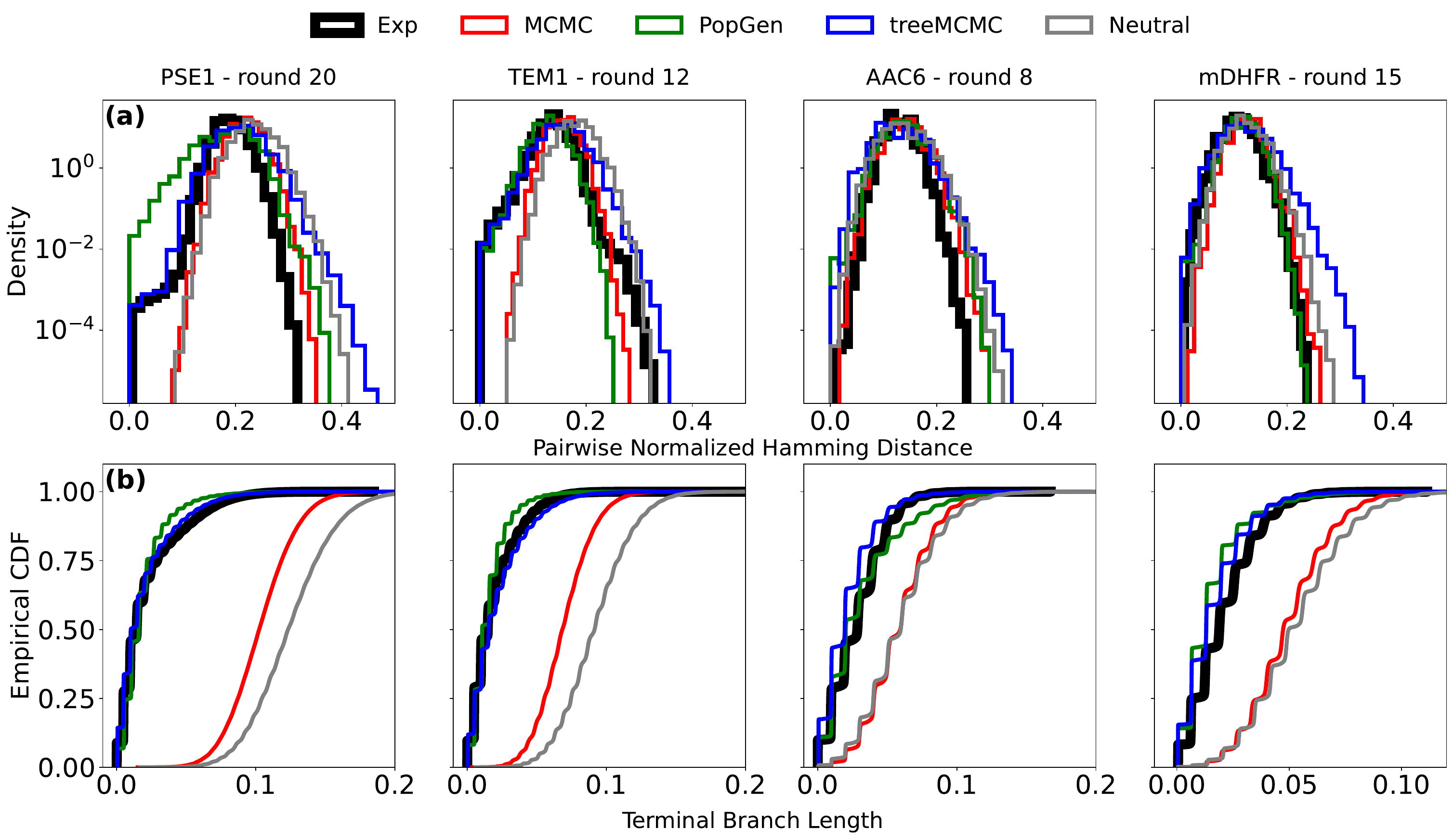}
    \caption{\textbf{Comparative Phylogenetic Statistics.} Distributions of evolutionary metrics for the PSE1, TEM1, AAC6, and mDHFR experiments against simulation results. The top row (a) displays the normalized pairwise Hamming distances between sequences, and the bottom row (b) illustrates the cumulative density function of terminal branch lengths derived from the inferred phylogenetic trees. Neutral (grey) curves corresponds to a PopGen simulation without the selection and amplification steps.}
    \label{fig::pairwise}
\end{figure*}

\subsection{Divergence \& Selection}
The initial phase of our analysis examines whether the computational models accurately reproduce the fundamental statistical profile of the evolved populations. As shown in Fig.~\ref{fig::tuning} for the PSE1 experiment, calibrating the simulations to match the observed mean Hamming distance from wild-type $\langle H \rangle$ and the average DCA energy $\langle E \rangle$ is sufficient to recover reasonably well the full experimental distributions of both metrics. 
This alignment demonstrates that all three computational frameworks—despite their different underlying assumptions—can effectively mirror the global characteristics of the experimental evolution process. 
The congruence in the energy distributions indicates that the DCA-inferred landscape, when coupled with our evolution protocols, successfully captures the fitness distribution and the selective pressure occurring in the laboratory. 
Similarly, the accurate reconstruction of the Hamming distance distributions suggests that the models correctly parameterize the mutational load and the evolutionary distance traversed from the ancestral sequence. 
From this high-level perspective, the results suggest that at relatively short evolutionary timescales, the specific nature of the simulation engine is secondary to the quality of the underlying fitness landscape. 
If the objective is simply to estimate the aggregate population fitness or the degree of sequence divergence, a standard MCMC approach using independent trajectories serves as a biologically reasonable approximation. 
This consistency is not isolated to PSE1 but is a robust feature across all four proteins analyzed (PSE1, TEM1, AAC6, and mDHFR; see Supplementary Materials).
However, these global metrics essentially treat the population as a collection of independent entities, masking the historical dependencies and lineage structure that define real biological evolution. 
While all models appear equivalent when looking at the ``snapshot'' of the final population, they differ fundamentally in how they navigate sequence space. 
In the following sections, we move beyond these aggregate statistics to evaluate the phylogenetic fidelity and internal diversity of simulations—metrics that reveal the true capacity of a framework to capture the non-equilibrium nature of the evolutionary process.

\begin{figure*}
    \centering
    \includegraphics[width=0.8\textwidth]{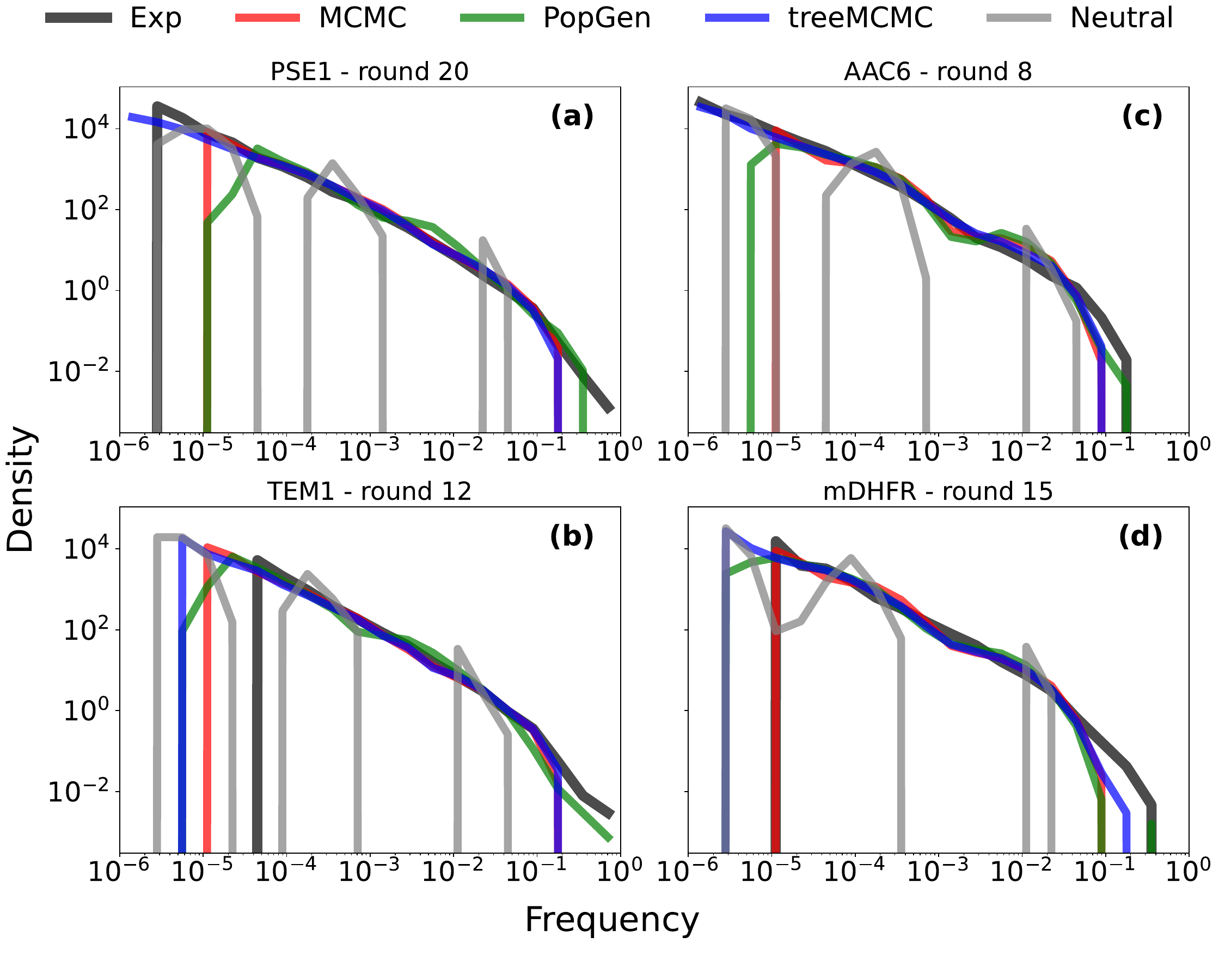}
    \caption{\textbf{Site Frequency Spectra (SFS)}. The frequency distribution of mutations is compared across the PSE1, TEM1, AAC6, and mDHFR experiments for all simulation schemes. Neutral (grey) simulation corresponds to a PopGen setup without a selection step. Curves are averaged over $10$ different realizations of each dynamics.}
    \label{fig::sfs}
\end{figure*}

\subsection{Diversity \& Phylogeny}
The capacity of a simulation to accurately reproduce the phylogenetic structure is a critical and robust test of its historical fidelity~\cite{neher2014predicting}. 
This structure, which captures the detailed history of divergence and common ancestry within the evolving population, is assessed here by studying the distribution of pairwise Hamming distances between all sequences and the distribution of terminal branch lengths in the phylogenetic trees.
The former can be directly computed from the synthetic sequences, while the latter, which quantifies the amount of evolutionary change accumulated by a specific lineage since diverging from its most recent common ancestor, is measured by inferring a phylogenetic tree from the simulated sequences, after removing duplicates. 
As illustrated in Fig.~\ref{fig::pairwise}, the standard MCMC scheme systematically fails to reproduce the detailed structure of these distributions across all four experiments (PSE1, TEM1, AAC6, mDHFR). 
This failure stems from the inherent nature of MCMC, which is designed to navigate the fitness landscape through independent Markov chains. 
This effectively washes out the historical dependencies and the signals of common ancestry that are crucial to the complex tree structure observed in the experimental data. This deficiency is particularly noticeable in the left tail of the pairwise distance distribution and in the absence of short terminal branch lengths, where the model underestimates the number of closely related sequences.

In stark contrast, both the treeMCMC and the PopGen schemes show strong agreement with the experimental distributions. 
This superior performance confirms that simply defining a fitness landscape is insufficient; the simulation must incorporate an explicit historical component to accurately model evolutionary path and structure. 
The treeMCMC framework achieves this by imposing a pre-inferred phylogenetic backbone onto the evolutionary dynamics. 
Its success in reproducing these quantities is not surprising, as the phylogenetic information of the experiments was encoded in the model.
On the other hand, the PopGen scheme naturally captures these traits by explicitly modeling the non-equilibrium processes of mutation, selection, and finite population sampling over discrete rounds, thereby generating a compatible phylogenetic structure as an emergent property of the dynamics. 
As a baseline, we also included in the analysis the case of neutral evolution, which in our framework corresponds to a PopGen scheme without the selection and amplification steps. 
For the sake of a correct comparison the mutation rate $\nu$ of the neutral simulation is still tuned to reproduce the mean divergence from the wild-type. 
It is quite clear that neutral simulations skipping the selection and amplification step do not correctly reproduce the statistics of experimental samples.
Obviously, sequences from neutral evolution are much less likely to be functional under the experimental selective pressure, as their DCA energy is higher with respect to the output of other simulations and it was showed in~\cite{russ2020evolution} that high model energy is an hallmark of non-functionality.

\subsection{Mutational Spectra}
The analysis of the mutational spectrum provides an essential, population-level view of the evolutionary dynamics, offering insight into the processes that shape genetic variation. 
Specifically, the Site Frequency Spectrum (SFS) is a highly informative summary statistic that quantifies the distribution of allele frequencies across all polymorphic sites in a population. In our case it is visualized by showing the  histogram of frequency of single amino acid mutations with respect to the wildtype. 
By differentiating between mutations that are rare, intermediate in frequency, or nearly fixed, the SFS serves as a sensitive measure of the balance between mutation, selection, and genetic drift. 
Strong positive selection, common in directed evolution, is expected to result in an over-representation of high-frequency variants (as beneficial mutations rapidly sweep to fixation) and a corresponding reduction in the number of low-frequency (rare) variants, which are quickly purged or favored. 
Therefore, accurately reproducing the experimental SFS shape is a stringent test of a model capacity to integrate these forces correctly.

\begin{figure*}
    \centering
    
    \includegraphics[width=\textwidth]{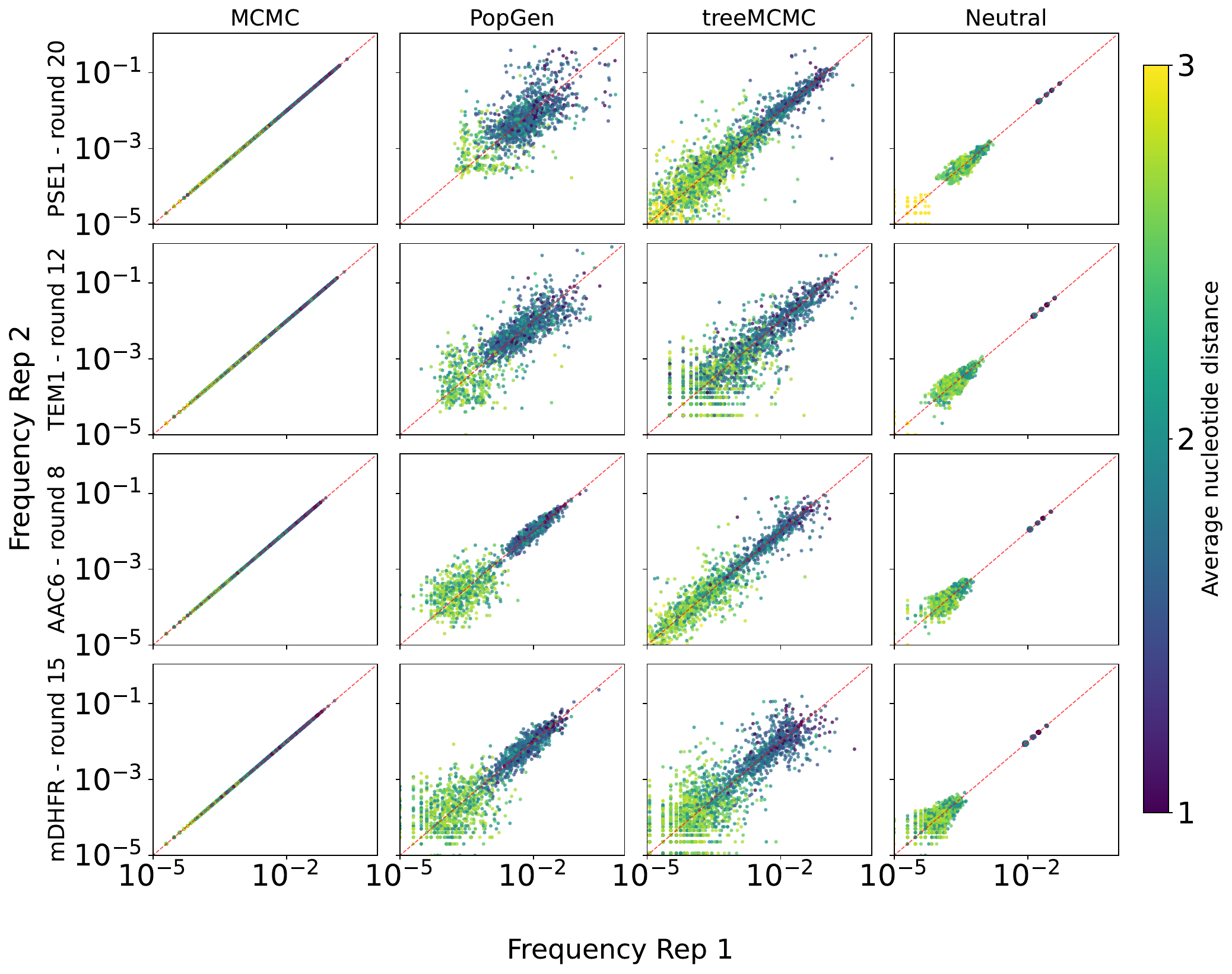}
    \caption{\textbf{Reproducibility of variant frequencies simulation methods.} 
    Each panel displays a log-log scatter plot comparing variant frequencies between two replicates. 
    Rows represent different experiments at specific selection rounds: PSE1 (round 20), TEM1 (round 12), AAC6 (round 8), and mDHFR (round 15). 
    Columns denote the modeling approach used: MCMC, PopGen, treeMCMC, and Neutral. 
    Points are colored according to the average nucleotide distance between the codons of the wildtype and of the mutated amino acid. 
    The red dashed line indicates the identity diagonal ($y=x$), representing perfect correlation between replicates.}
    \label{fig:replicate_main}
\end{figure*}

Fig.~\ref{fig::sfs} illustrates the SFS for the four experiments with all the different models. 
The distribution of mutation frequencies is compared across the simulation schemes against the experimental data. 
The SFS plots are averaged over $10$ different realizations. 
The neutral simulation (gray), which corresponds to a PopGen simulation without selection and amplification steps, shows the expected distribution for a purely neutral process, with a severe under-representation of high-frequency alleles, confirming that selection is the dominant factor shaping the experimental spectrum. 
Standard MCMC (red) and treeMCMC (blue) schemes successfully shift the distribution away from neutrality by incorporating fitness selection. 
Both MCMC approaches systematically underestimate the count of high-frequency variants (those above $0.1$), which represents the signature of episodic fixation events driven by positive selection, and show deviations in the intermediate frequency bins. 
The PopGen scheme shows a better agreement in terms of emergence of high-variant mutant, while losing accuracy in reproducing low frequency mutations. 

One should note the discontinuity of the neutral SFS, which is characterized by a big bulk at low frequency and a smaller peak over intermediate ones. 
If one imagines to run the neutral simulation up to equilibration, all frequencies would be equally represented as $1/q$ where $q=20$ is the total number of amino-acids. 
Yet, the SFS is influenced by the accessibility map of the wild-type, which constrains evolution at short time scales. 

\begin{figure*}
    \centering
    \includegraphics[width=0.8\textwidth]{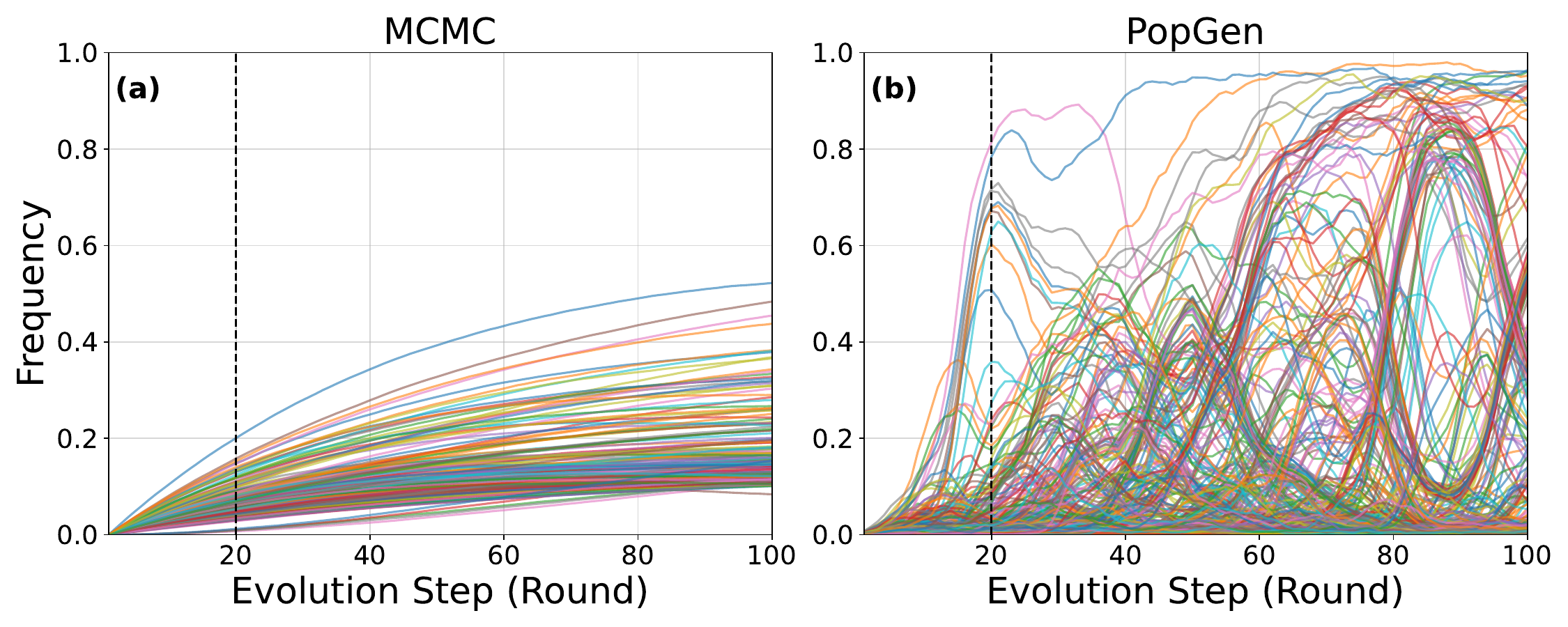}
    \caption{\textbf{Mutational dynamics.} Evolutionary trajectories of individual mutations that emerge and sweep through the population, as predicted by (a) MCMC simulations and (b) PopGen simulations. Mutation frequency is plotted against the equivalent of experimental evolution rounds. Parameters are set to reproduce the PSE1 experiment at round $20$. Only mutations whose frequency starts below $0.1$ and reaches a maximum frequency greater than or equal to $0.1$ are shown, demonstrating true emerging beneficial or neutral mutations.}
    \label{fig::mut_dyn}
\end{figure*}

In Fig.~\ref{fig:replicate_main} we compare the mutant frequencies of two replicate simulations for all methods and experiments, colored by their average nucleotide distance from the wild-type codon. 
We observe that in neutral simulations only very close amino acids are sampled at intermediate abundances, with samples being trapped in the neighborhood of the initial sequence.  
It is important to then address the fact that the high correlation between two neutral (and hence completely random) simulations is totally caused by the accessibility bias. 
Interestingly, this accessibility effect, with replicates being more correlated at high frequencies, is also quite dominant in the PopGen scheme. 
In addition, the replicate analysis shows that standard MCMC replicates are perfectly correlated, whereas treeMCMC and PopGen display more noise between different realizations. 
Unfortunately, to the best of our knowledge, it has never been studied how two different experiments with the same conditions correlate, but we must expect that some noise should be present, corroborating our hypothesis that treeMCMC and PopGen are more suitable for reproducing experimental results.

\subsection{Mutational Dynamics}

While the site frequency spectrum summarizes mutational prevalence at a static endpoint, evaluating evolutionary models requires scrutinizing the dynamical paths that mutations follow to reach those frequencies. We analyzed the temporal trajectories of single mutations as a critical test for the non-equilibrium behavior of MCMC and PopGen schemes. 
We define emerging trajectories as mutations transitioning from an initial frequency below 0.1 to a maximum observed frequency exceeding 0.1. This filtering isolates the dynamics of beneficial or quasi-neutral mutations driving adaptation while excluding transient background variation. Fig.~\ref{fig::mut_dyn} compares these emerging trajectories for the PSE1 protein. Trajectories generated by the MCMC model exhibit gradual, nearly monotonic increases. While the MCMC process identifies high-fitness regions via stochastic sequence-level sampling and Boltzmann-weighted acceptance, it lacks kinetic realism. The absence of sharp, episodic increases suggests that the sequential, single-sequence sampling inherent to MCMC chains dampens the abrupt, population-scale dynamics characteristic of real selective sweeps.
In contrast, the PopGen scheme produces abrupt, episodic frequency changes that closely resemble classical selective sweeps. 
By integrating explicit population-level forces and generational turnover, this model demonstrates greater fidelity to the non-equilibrium dynamics observed in real-world evolution experiments. 
This dynamic analysis confirms that while the MCMC scheme provides accurate equilibrium estimates of the fitness landscape, its intrinsic dynamics fail to capture the mechanics of population-driven sweeps. 
We expect that if experiments were extended with sufficient statistics, the observed curves would align more closely with PopGen predictions. 
Developing these simulations allows us to go beyond current experimental timescales, serving a predictive tools to guide future \textit{in vitro} experimental evolution strategies by identifying high-probability mutational pathways and optimizing selection pressures before laboratory implementation.

\subsection{Long-time behavior and approach to equilibrium}

\begin{figure*}
    \centering
    \includegraphics[width=0.7\textwidth]{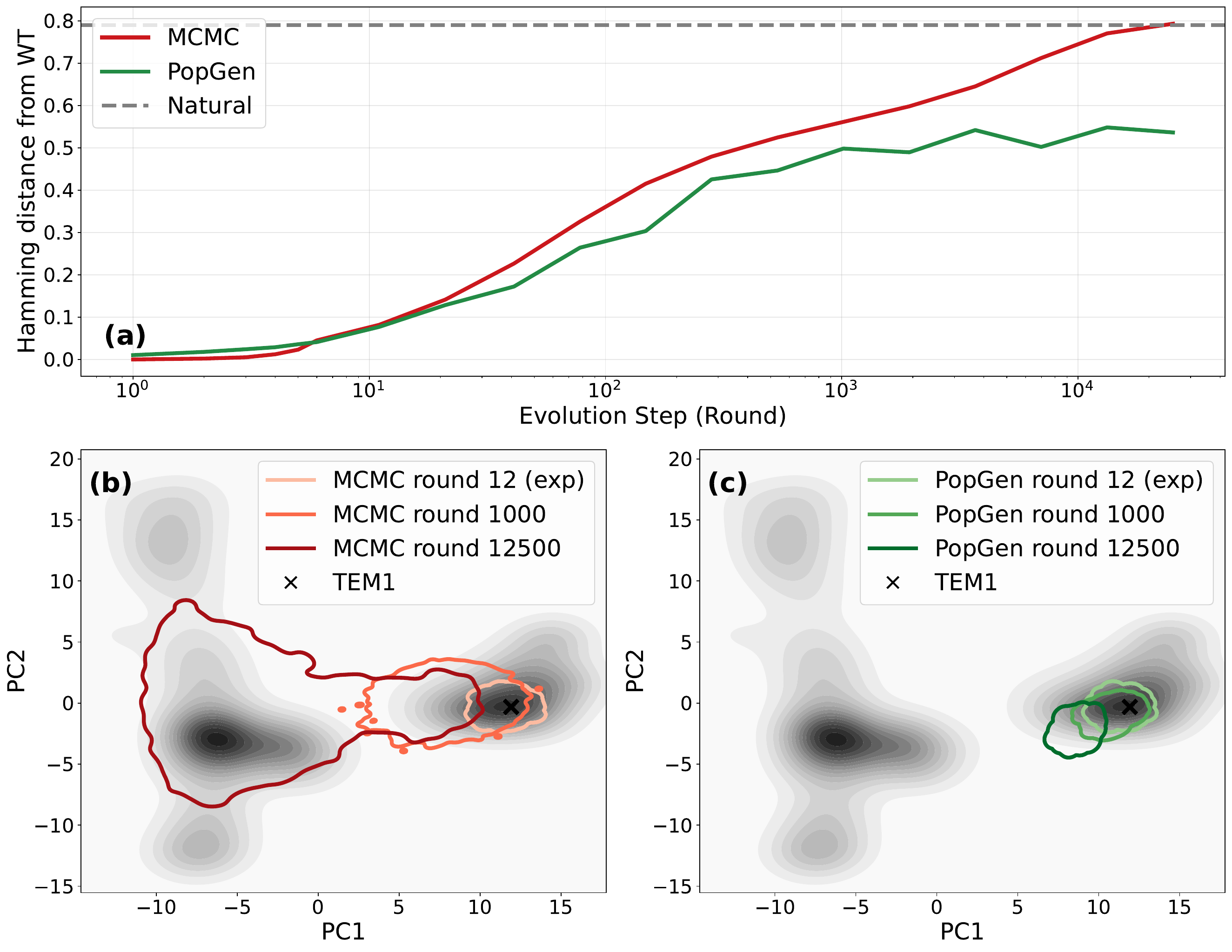}
    \caption{\textbf{Long term behavior.}
    (a) Average Hamming distance from the wild-type, $\langle H \rangle$, as a function of evolutionary time $t$ (in units of experimental rounds) for MCMC and PopGen simulations. (b) Projection of MCMC-generated sequences onto the principal components (PCs) of the natural MSA at different evolutionary times. (c) Same as (b) for PopGen simulations.}
    \label{fig::long_term}
\end{figure*}

In this work, we focus on how accurately algorithms can reproduce experimental results, in order to use them as a predictive tool to inform future work. 
Laboratory time scales are relatively short, although promising recent advances in experimental techniques 
may allow to reach much longer times~\cite{rix2023continuous}.
Here, we can however speculate on the long-time limit by looking at the different models we described. 
By definition, for a sufficiently large evolutionary time, the MCMC, which respect detailed balance, will produce sequences that correspond to independent samples of the distribution $P(\boldsymbol{a})$ in Eq.~\eqref{eq:prob}.
Similarly, the treeMCMC algorithms will reach the same point in the limit
of very large mutation rate $\mu$.
The situation is much less clear for the PopGen scheme, as in this case there is no detailed balance and equilibrium is not reached.
In Fig.~\ref{fig::long_term} we show how the models perform when we let the simulation continue for multiple rounds.
In panel (a) we plot the average Hamming distance of the evolving sequences from the wild-type, $\langle H \rangle$, as a function of the evolutionary time $t$ for MCMC and PopGen. 
In MCMC, the Hamming distance saturates at a value compatible with the average Hamming distance between natural sequences, indicating that the model reached equilibrium. 
The PopGen curve instead saturates at a much smaller value, as the evolving sequences seem to be stuck in a region around the original wild type.
This is confirmed in the bottom panels. 
In panel (b) we show how the synthetic sequences obtained via MCMC occupy the space of the Principal Components (PC) of natural sequences during evolution.
We see that at short evolutionary times the sequences are confined around the wild-type, while as time goes by they spread and occupy a vast region of the space. 
Notice that the synthetic sequences do not cover exactly the distribution of the natural sequences. 
This is expected, as the sequences obtained here are sampled with $T\neq 1$ (calibrated as above) and without introducing or removing gaps. 
In the standard case with $T=1$ and evolvable gaps, the equilibrium sequences reproduce quite well the statistics of the natural MSA~\cite{DiBari2024}.
The situation is completely different in panel (c), where we plot instead PopGen sequences.
Here, as the evolutionary time increases, the sequences move around a little bit with respect to the position of the wild-type, but they never spread enough to cover a larger amount of the space.

Notice that commenting more on the implications of these results for long-term natural evolution is hard. 
While we can argue that natural sequences represent a sample of proteins with a specific function, these sequences have been produced by a complex, very long evolutionary process that is very different from the {\it in vitro} controlled evolution modeled here.

\section{Discussion}

In this work, we performed a rigorous comparative analysis of three computational schemes---standard Markov Chain Monte Carlo (MCMC), MCMC on a phylogenetic tree (treeMCMC), and a population genetics (PopGen) model---for simulating the non-equilibrium process of protein evolution on fitness landscapes defined by Direct Coupling Analysis (DCA). The core scientific question addressed is whether generative models that are highly successful at reproducing the statistical properties of a sequence ensemble at equilibrium can also accurately capture the evolutionary paths, kinetics, and population structure observed in laboratory evolution experiments. It should be noted that while most of the results we obtained are consistent with expectations based on simple, stylized models, here we are addressing this question in a realistic, quantitative context, where we consider actual experimental data on real proteins, and the underlying fitness landscape is also inferred from data.

Our investigation spanned two crucial dimensions of evolutionary modeling, benchmarked across four distinct experimental datasets (PSE1, TEM1, AAC6, and mDHFR). First, we evaluated the ability of each scheme to preserve the correct phylogenetic structure of the evolving sequence population while maintaining the same divergence from the wild-type and selective pressure observed in experiments. We found that the standard MCMC scheme, operating under the assumption of independent sampling from the Boltzmann distribution, is inherently limited by its independent chain nature and fails to reproduce the essential features of a branching evolutionary history. The treeMCMC scheme offered a significant methodological improvement, successfully incorporating the historical dependencies by aligning evolution with a predefined genealogical structure. However, the PopGen scheme, by explicitly modeling mutation, selection, and drift within a growing population, proved to be a robust predictor without the need of inferring additional data from experiments, reproducing experimental patterns as emergent properties of the simulation.

The second, and perhaps most critical, validation came from analyzing the mutational dynamics and population-level features. While the Site Frequency Spectrum was reasonably well-approximated by all models, the true divergence in kinetic behavior became apparent when examining individual emerging trajectories (Fig.~\ref{fig::mut_dyn}). The standard MCMC process yields trajectories characterized by a gradual and nearly monotonic increase in frequency, reflecting the dampened dynamics of sequential, single-sequence sampling. This kinetic profile is demonstrably inconsistent with the rapid, episodic increases associated with strong selective sweeps in a real population.
In sharp contrast, the PopGen model generated trajectories that closely mimic the theoretical curves of classical selective sweeps, characterized by a smooth, accelerated increase in frequency followed by eventual fixation. This observation validates the superior realism of the PopGen scheme, confirming that explicitly modeling population size, mutation rate, and generational turnover is essential for correctly capturing non-equilibrium dynamics.

A primary drawback of the PopGen approach is, however, that its dynamics do not guarantee convergence to the underlying probability distribution defining the fitness landscape. When simulated over long evolutionary time scales, sequences produced by this scheme tend to remain trapped in the proximity of the wild-type. While one might expect the variability seen in natural sequences to eventually be recovered, as a single MCMC chain would eventually drift toward equilibrium distances for purely entropic reasons, the PopGen model behaves differently. 
This localization may be explained by the large population size effectively acting as a lower \textit{effective temperature} (i.e. higher selective pressure), or by the fact that when time is large compared to population size, all individuals eventually descend from a single most recent common ancestor, causing the population to behave as one effective trajectory rather than an independent ensemble. Indeed, experimental evolution ignores critical ingredients such as time-varying selection pressures, complexity of genotype-phenotype mapping, and spatial population structures that could contribute to the emergence of natural variability.
Beyond temporal shifts, niche-dependent selection may also play a critical role in explaining the existence of multiple fitness peaks in natural landscapes. Our current framework's failure to sample the full breadth of these peaks could stem from the assumption of a uniform environment; in reality, different subtrees within a global phylogeny may experience distinct selective pressures, allowing the population to explore and occupy disparate regions of the fitness landscape simultaneously. Advanced simulation schemes incorporating these time- and niche-dependent environments may clarify these roles in future work.

Despite these limitations at extreme scales, the validated PopGen model provides a crucial conceptual and predictive advantage for interpreting \textit{in vitro} evolution. 
Real-world experiments are often restricted by technical constraints; they rarely sequence every intermediate round or extend long enough to capture full trajectories up to fixation under high-statistics conditions. 
This framework allows for the reliable extrapolation of dynamics beyond these experimental limits. 
It offers a rigorous, prospective view of mutational dynamics, suggesting that if experiments were conducted with sufficient statistical power over extended periods, the observed trajectories would align with the smooth, rapid sweeps predicted by the PopGen model.

Ultimately, our understanding of these evolutionary dynamics can be divided into three distinct time scales:
\begin{itemize}
    \item Short (Experimental) Scales ($\sim 10-20$ rounds): Models appear to capture the essential details of this regime, aligning well with available experimental results.
    \item Intermediate Scales ($\sim 500-1000$ rounds): Although experimental data is currently limited, technological advances are expected to fill this gap shortly. In this regime, models suggest that the influence of epistasis becomes more pronounced~\cite{DiBari2024,rossi2024fluctuations}.
    \item Long (Natural) Scales ($\sim 10^6$ rounds): At this limit, the MCMC scheme successfully reproduces natural sequence variability, whereas the PopGen approach remains localized near the starting sequence. The outcome of an {\it in vitro} experiment sustained over such time scales remains an open question.
\end{itemize}

This work lays a rigorous foundation for several promising avenues of future research aimed at further elucidating the evolutionary process on complex fitness landscapes as the PopGen model could be used as a predictive tool to study complex biological phenomena. Furthermore, integrating the validated PopGen framework with machine learning or reinforcement learning approaches could lead to the development of sophisticated algorithms capable of predicting the optimal sequence of beneficial mutations and the expected time-to-fixation for a given fitness target, thereby directly impacting and streamlining future {\it in vitro} evolution protocols. 
Finally, given the computational demands of the PopGen simulation due to the lack of parallelization, future efforts should focus on identifying and integrating simplifying assumptions to achieve robust and scalable simulations at a reduced computational cost.

\section*{Acknowledgements}
We thank Martin Weigt for insightful discussions. 
This research has been supported by first FIS (Italian Science Fund) 2021 funding scheme (FIS783 - SMaC - Statistical Mechanics and Complexity) from MUR, Italian Ministry of University and Research and from the PRIN funding scheme (2022LMHTET - Complexity, disorder and fluctuations: spin glass physics and beyond) from MUR, Italian Ministry of University and Research. AMW and TM were supported by the Foundation pour la Recherche Medicale grant EQU202503019997.

\bibliography{references_nourl}

\clearpage

\onecolumngrid

\begin{center}
    \vspace*{1cm}
    {\large \bf Supplemental Material for \\ 
    ``Modeling Protein Evolution via Generative Inference From Monte Carlo Chains to Population Genetics''}
    \vspace{0.5cm}
\end{center}


\renewcommand{\thesection}{S\arabic{section}}
\renewcommand{\thefigure}{S\arabic{figure}}
\renewcommand{\thetable}{S\arabic{table}}
\setcounter{section}{0}
\setcounter{figure}{0}
\setcounter{table}{0}
\setcounter{page}{1}

\section{Detailed description of the MCMC simulation scheme}
\label{suppsec:MCMC_details}

To accurately model protein evolution as a stochastic process, we define our evolutionary framework at the nucleotide level as a sequence $\mathbf{c} = (c_1, \dots, c_L)$ of codons (with $c_i = (n_{i,1}, n_{i,2}, n_{i,3}) $ being the codon in the $i^{th}$ position), which is translated to the amino acid sequence $\mathbf{a}(\mathbf{c})$ via the genetic code. 
This approach enables the explicit simulation of molecular events like single-nucleotide mutations and codon-level insertions/deletions (indels). 
To generate a population of sequences, one runs independent Markov Chain Monte Carlo chains which start from the same wild-type sequence of interest. 

The model uses a nucleotide sequence energy $E(\mathbf{c})$ to guide the dynamics and enforce detailed balance. This energy extends the DCA-derived amino acid energy $E[\mathbf{a}(\mathbf{c})]$ (inverse fitness) by integrating codon bias $f(c_i | a_i)$:
$$E(\mathbf{c}) = E[\mathbf{a}(\mathbf{c})] - T \sum_{i=1}^{L} \log f(c_i | a_i).$$
As discussed in the main text, here we assume no codon bias and we assign equal probability to all the codons that code for amino acid $a_i$, so that $f(c_i \mid a_i) = 1/N_{a_i}$, with $N_{a_i}$ the number of codons coding for a certain amino acid $a_i$.

The evolutionary dynamics is driven by a mixed MCMC sampler that combines two types of moves, namely single-nucleotide mutations and codon insertions or deletions, in an iterative process:
\begin{enumerate}
    \item Single-Nucleotide Mutations  (with probability $1 - p$)
    \begin{itemize}
        \item Mechanism: Gibbs Move.
        \item Action: A non-gapped codon position $i\in \{1,\dots,L\}$ and nucleotide position $k\in\{1,2,3\}$ are chosen uniformly at random. 
        The nucleotide at that position is updated by sampling a new nucleotide from the conditional distribution
        \begin{equation}
        P(n_{i,k} = n \mid \mathbf{n}_{\neq i}) 
        = \frac{e^{-\beta E(n_{1,1}, n_{1,2}, \dots, n_{i,k}=n, \dots, n_{L,3})}}
        {\sum\limits_{n' \in \mathcal{N}} e^{-\beta E(n_{1,1}, n_{1,2}, \dots, n_{i,k}=n', \dots, n_{L,3})}},
        \end{equation}
        where $\mathcal{N}$ denotes the nucleotide alphabet and $E(\mathbf{n})$ is the nucleotide-level energy function. 
        $\beta = 1/T$ is the inverse effective temperature, which serves as a proxy for selective pressure, and which is tuned in order to better reproduce the experimental results.
    \end{itemize}
    \item Codon Insertions/Deletions (with probability $p$)
    \begin{itemize}
        \item Mechanism: Metropolis Move.
        \item Action: A site $i$ is chosen uniformly and a move from the current codon $c_i$ to a new one $c'_i \in \{ (---), (AAA), (AAC), \dots, (TTT)\}$ is proposed.
        As the single nucleotide mutations are already considered in the previous move type, here we only consider moves that: change a gapped site into a non-gapped site, change a non-gapped site into a gapped site, or do not change the current codon.
        These transition probabilities are summarized in the transition matrix $\Pi$, defined as a symmetric matrix with zeros at transitions between different non-gapped codons:
        \begin{equation}
        \begin{array}{c|cccc}
        \Pi & --- & AAA & AAC & \cdots \; TTT \\ \hline
        ---   & \eta & \kappa & \kappa & \cdots \; \kappa \\
        AAA  & \kappa & \gamma & 0 & \\
        AAC   & \kappa & 0 & \gamma & \\
        \vdots& \vdots &  &  & \ddots \\
        TTT & \kappa &  &  & \gamma
        \end{array}.
        \label{eq:proposal_matrix}
        \end{equation}
        The transition matrix $\Pi$ allows insertion and deletion events, but disallows direct codon-to-codon substitutions. 
        The parameters $\eta$ and $\gamma$ ensure proper normalization of the proposal probabilities and correspond to null moves that leave the codon unchanged. To accelerate the sampling, we choose $\kappa$ as large as possible while preserving normalization. This leads to $\eta = 0$, $\kappa = 1/64$, and $\gamma = 63/64$. Consequently, when an amino-acid–coding codon is selected, the proposal leaves the codon unchanged in $63$ out of $64$ cases, and proposes a triple-gap insertion only in a single case. Proposals leading to stop codons ($c'_i \in \{ (TAA), (TAG), (TGA) \}$) are systematically rejected.
        \item Acceptance: The move is accepted based on the energy difference, ensuring detailed balance 
        $$p(c_i \to c'_i) = \min \left\{ 1, e^{-[E(\mathbf{c}') - E(\mathbf{c})]/T} \right\} $$
    \end{itemize}
\end{enumerate}
Both move types enforce detailed balance, guaranteeing that the entire algorithm converges to the inferred probability distribution of natural sequences.
In this work, we choose $p=0$, ignoring indel (insertion/deletion) mutations, as in~\cite{DiBari2024}, as we want to focus mainly on short time scales.

\clearpage

\section{Detailed description of the PopGen simulation scheme}
\label{suppsec:Popgen_details}

The PopGen simulation scheme follows the same idea of~\cite{barton2016relative,Hart_2019}.
It was also previously used in the context of landscape inference from directed evolution experiments~\cite{Sesta2021amala}.
This framework explicitly models the core evolutionary processes over discrete rounds, providing a more realistic, though computationally intensive, representation of directed evolution. The simulation maintains a fixed total population size ($N_{\rm tot}$) and evolves via Wright-Fisher-like dynamics, where the interplay of stochastic mutation and selection captures the species-competition nature of the laboratory setting. Starting from a set of replicates of the common ancestral sequence, the simulation iterates evolutionary rounds comprising the following steps:

\begin{enumerate}
    \item Mutation: sequences in the current population are mutated in non-gapped positions according to a user-defined single site mutation probability $\tilde{p}$. This probability is obtained from the single site mutation rate ($\nu$) of a generalized Jukes-Cantor model, where (differently from previous applications~\cite{Sesta2021amala} that worked in amino-acid space) $q=4$ is the number of nucleotides. 
    We get
    $$\tilde{p} = \frac{q - 1}{q} \left( 1 - e^{-\nu t} \right) \quad (\text{with } t=1).$$
    In this case as well, stop codons are always avoided, so that only mutations to viable codons are proposed.
    Notice that in principle it would be possible to add indel mutations to this setting, as briefly discussed in the main text, but we defer it to future work as it is not needed here.
    \item Selection: newly generated sequences are then selected based on their fitness. 
    The survival probability for a sequence $\boldsymbol{a}$ is $$W(\boldsymbol{a}) = \frac{1}{1+e^{\tilde{\beta} \left( E(\boldsymbol{a})-\tilde{E} \right)}}$$   where $E(\boldsymbol{a})$ is the energy defined by the DCA model. 
    Additionally, $\tilde{E}$ acts as a selective threshold: if a sequence $\boldsymbol{a}$ has a DCA energy above the threshold, its probability of surviving decays exponentially. Finally, $\tilde{\beta}$ is the selection strength that determines the scale of the exponential decay in survival probability for high energy sequences.
    \item Amplification: the total population size is maintained via multinomial sampling from the selected sequences, completing the evolutionary cycle.
\end{enumerate}

\clearpage

\section{Calibration procedure}
\label{suppsec:calibration_details}

\begin{figure}[ht!]
    \centering
    \includegraphics[width=0.8\textwidth]{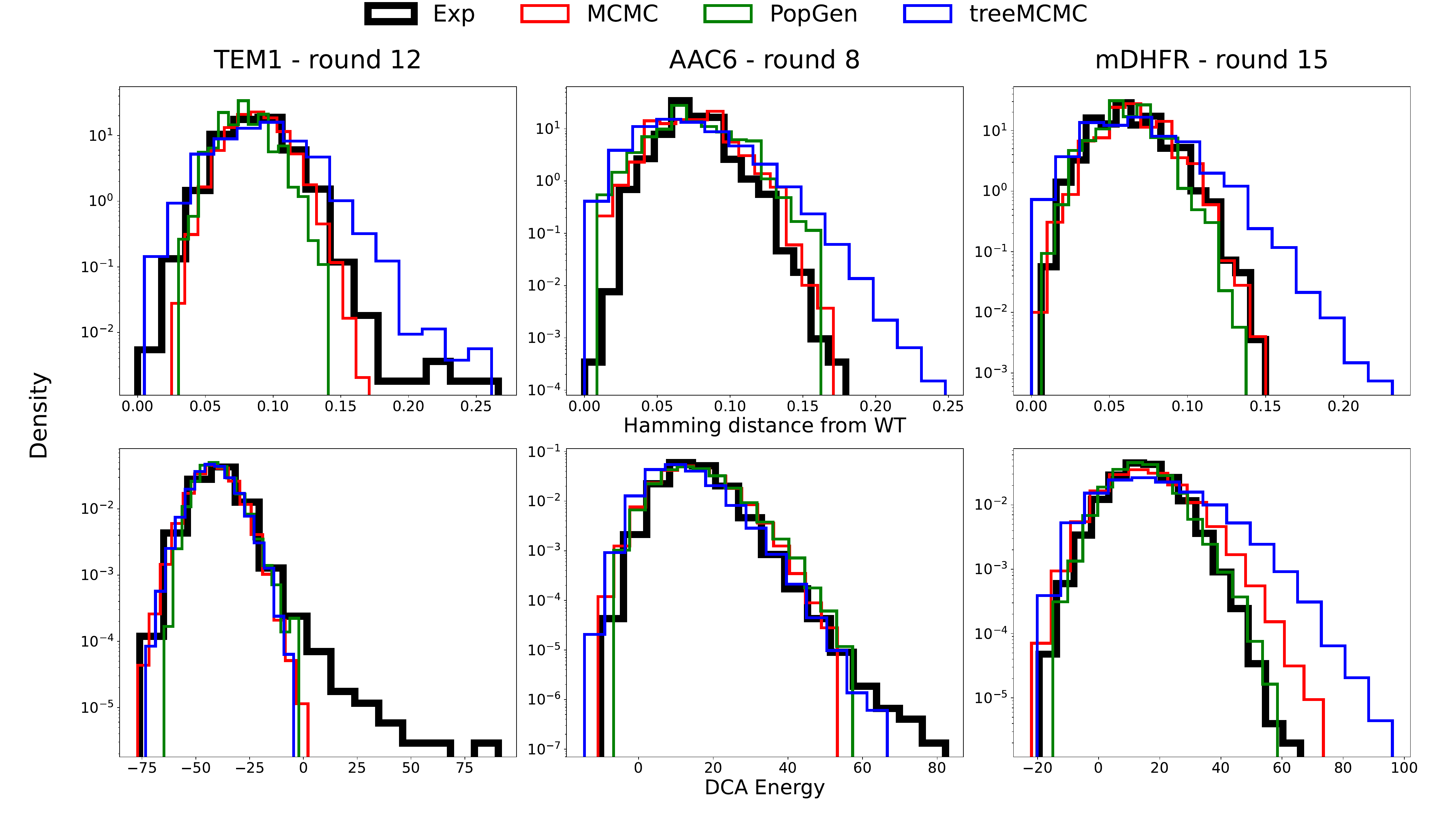}
    \caption{\textbf{Comparison of simulation frameworks across three directed evolution experiments.} (Top Row) Normalized Hamming distance from the wild-type sequence. (Bottom Row) DCA Energy distributions. Columns from left to right represent TEM1 (round 12), AAC6 (round 8), and mDHFR (round 15). Experimental data (black) is compared against in silico MCMC (red), PopGen (green), and treeMCMC (blue) simulations conducted with optimal parameters obtained with gridsearch.}
    \label{fig:supp_calibration}
\end{figure}

The models described in the main text and in the sections above are calibrated to reproduce two key observables of the experimental data: the average Hamming distance of evolved sequences from the starting sequence and the average DCA energy of the evolved sequences. 
In summary, the parameters adjusted during this calibration are $N_{\rm steps}$ and $\beta$ for MCMC, $\mu$ and $\beta$ for treeMCMC, and $\tilde{\beta}$, $\nu$, and $\tilde{E}$ for PopGen, with the number of rounds fixed to the experimental value for the latter. 
The specific parameter values were obtained via a grid-search procedure and are reported in Table~\ref{tab:S1} for each experiment and modeling framework.
We find that the number of MCMC steps required to reproduce the experimental observables is very small compared to the evolutionary time scales that the algorithm is capable of exploring~\cite{DiBari2024,rossi2024fluctuations}. 
The fine-tuned temperature $T = 1/\beta$ is only slightly higher than the training temperature $T = 1$ for PSE1 and TEM1, and noticeably higher for AAC6 and mDHFR. 
The selective pressure $\tilde{\beta}$ inversely correlates with the width of the energy distribution. In addition, when $\tilde{\beta}$ is significantly lowered, it effectively shifts the survival probability toward higher values. As a consequence, the selective threshold $\tilde{E}$ in the PopGen algorithm 
assumes values slightly lower than the typical DCA energies of experimental sequences, which generally remain near the energy of the starting sequence, consistent with the exploration of the local neutral space.

In Fig.~3 of the main text, we showed the comparison between experimental and simulated distributions of Hamming distances and DCA energies for the PSE1 wild-type. Here, in Fig.~\ref{fig:supp_calibration} we present the corresponding calibration results for the other cases (TEM1, AAC6, and mDHFR). 
With the exception of TEM1, which exhibits extended tails in the DCA energy distribution, the remaining systems show good agreement between in silico and experimental sequences.

\begin{table}[ht!]
\centering
\renewcommand{\arraystretch}{1.5}
\begin{tabular}{|l|cc|cc|ccc|}
\hline
& \multicolumn{2}{c|}{\textbf{MCMC}} & \multicolumn{2}{c|}{\textbf{treeMCMC}} & \multicolumn{3}{c|}{\textbf{PopGen}} \\
\cline{2-8}
Experiment & $N_{\rm steps}$ & $\beta^{-1}$ & $\mu$ & $\beta^{-1}$ & $\tilde{\beta}^{-1}$ & $\nu$ & $\tilde{E}$ \\
\hline
PSE1 - round 20  & 120 & 1.3 & 3.33 & 1.4 & 13.0 & 0.0085 & -55.0 \\
\hline
TEM1 - round 12  & 95  & 1.1 & 5.26 & 1.1 & 7.0  & 0.009 & -50.0 \\
\hline
AAC6 - round 8   & 30  & 2.1 & 2.09  & 1.8 & 10.0 & 0.012 & 0.0   \\
\hline
mDHFR - round 15 & 50  & 2.3 & 2.78  & 2.6 & 8.0  & 0.007 & 10.0  \\
\hline
\end{tabular}
\caption{Optimal simulation parameters to reproduce the four protein experiment across the three modeling frameworks. Values have been obtained by gridsearch to reproduce average Hamming distance from wild-type and average DCA energy of experimental sequences.}
\label{tab:S1}
\end{table}

\clearpage

\section{Intermediate rounds}
\label{suppsec:Intermediate_rounds}

\begin{figure}[t]
    \centering
    \includegraphics[width=0.4\columnwidth]{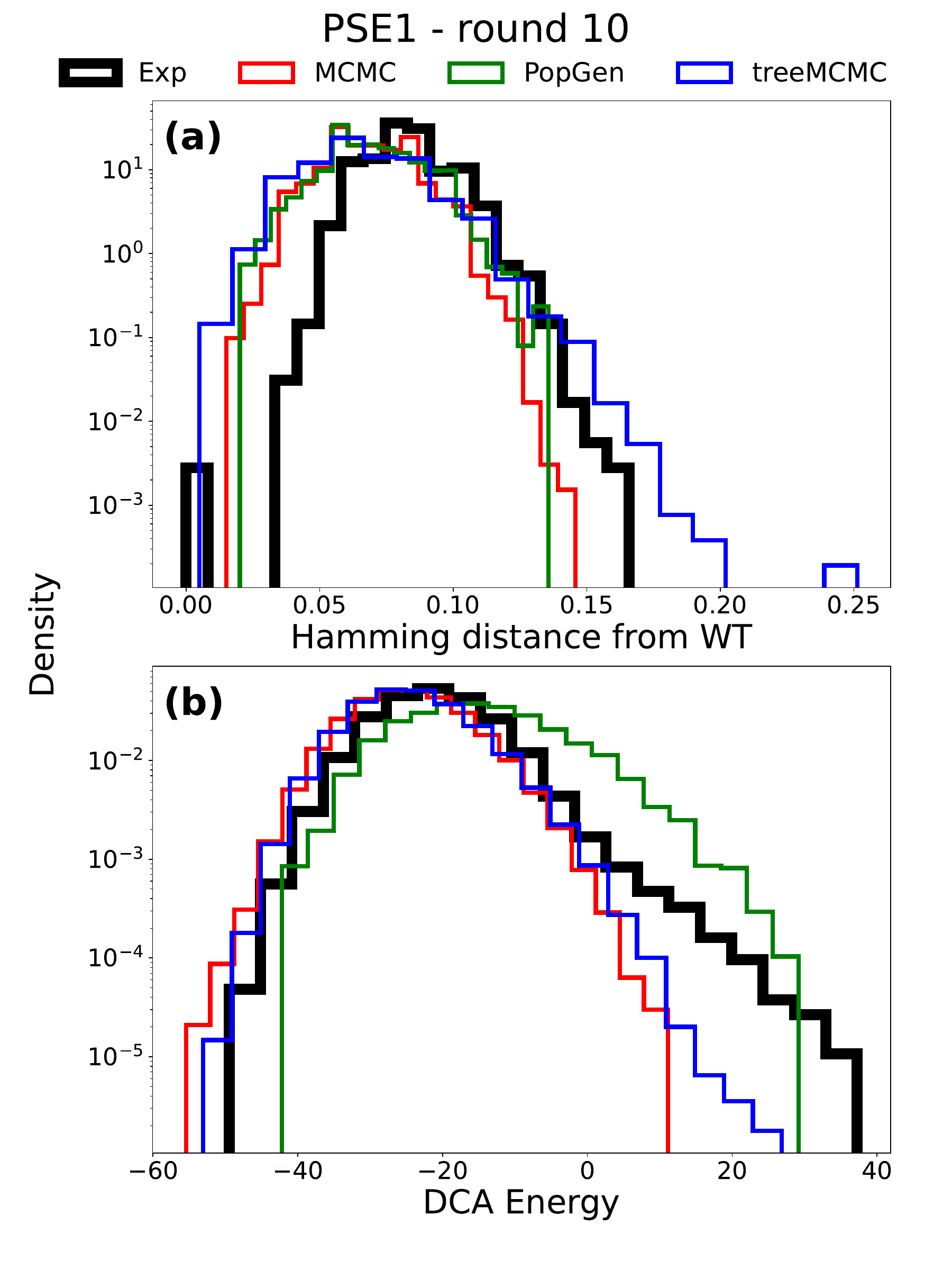}
    \caption{\textbf{Reproducing intermediate rounds} The distribution of (a) fractional Hamming distance $H/L$ from the wild-type and (b) DCA energy $E$ for the experimental PSE1 round $10$ population compared to the three simulation schemes. Here parameters are the same tuned to fit the last experimental round, with steps/rounds being rescaled linearly.
    }
    \label{fig::supp_tuning}
\end{figure}

Some of the experimental datasets analyzed in this work also include protein sequences obtained after fewer rounds of selection–mutation than the final round. 
For instance, in Ref.~\cite{stiffler2020protein}, the experimental sequences used for comparison with our in silico results in the main text correspond to round 20. 
To assess how well the models capture earlier stages of the experiment, we additionally compare them against sequences obtained after 10 rounds. Synthetic sequences generated with MCMC, treeMCMC, and PopGen are produced using models calibrated on the round-20 experimental data. To match the reduced evolutionary depth of the round-10 data, we halved the parameters 
that dictate the total simulation time in each algorithm: $N_{\rm steps}$ for MCMC, $\mu$ for treeMCMC, and the number of selection rounds for PopGen. The results are shown in Fig.~\ref{fig::supp_tuning}. Despite being fine-tuned on the final available round, all models are still able to reproduce the experimental distributions of Hamming distances and DCA energies with good accuracy.

\clearpage
\section{Consequences of codon accessibility}

\begin{figure}[ht!]
    \centering
    \includegraphics[width=0.7\textwidth]{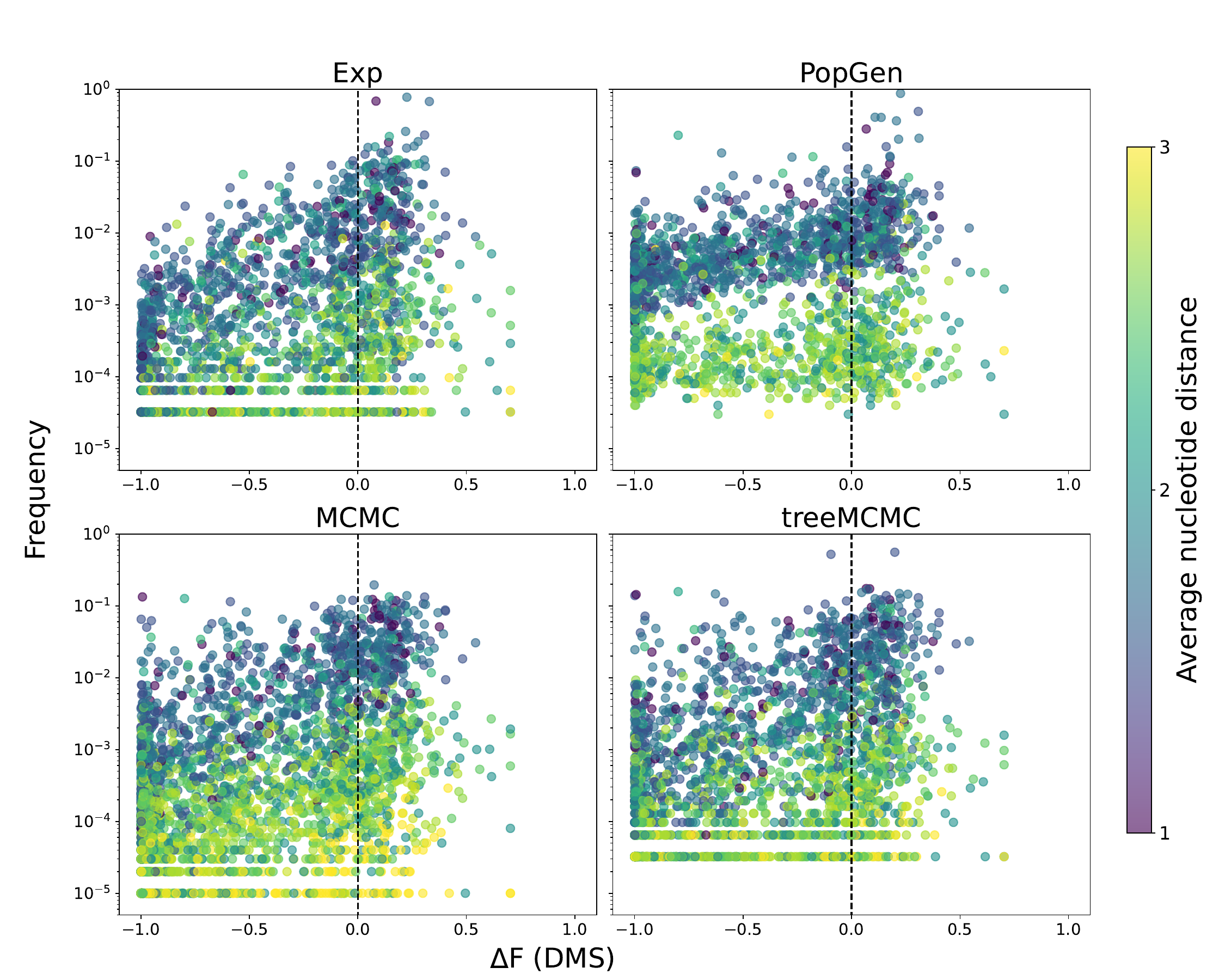}
    \caption{\textbf{Correlation between experimental fitness ($\Delta F$) and variant frequencies across modeling frameworks for TEM1 - round $12$.} Each panel shows the relationship between variant frequencies and deep mutational scanning (DMS) fitness measurements ($\Delta F$). Data points are colored according to their average nucleotide distance from the wild-type sequence. The four panels represent: (Top Left) Experimental frequencies; (Top Right) PopGen; (Bottom Left) MCMC; (Bottom Right) treeMCMC. All simulation frameworks were parameterized using the optimized values listed in Table S1. The $\Delta F$ values where obtained through deep mutational scanning~\cite{GONZALEZ2019fitness} of TEM1 variant. The dashed line is separating beneficial and deleterious mutations.}
    \label{fig:supp_codon_accessibility}
\end{figure}

As shown in~\cite{gunnarsson2023predicting}, the accessibility of a given amino acid from the wild-type genetic sequence can strongly affect its probability of being observed during short-time evolution experiments. 
The impact of this effect is illustrated in panel (b) of Fig.~2 of the main text, where the experimental frequency of each mutation at the final round of evolution is plotted against its fitness effect, as measured by deep mutational scanning (DMS). 
In Fig.~\ref{fig:supp_codon_accessibility}, we show the same comparison, plotting the DMS fitness change $\Delta F$ reported in Ref.~\cite{GONZALEZ2019fitness} against the frequency with which each mutation appears in synthetic sequences generated by the different algorithms. 
Compared to the experimental data shown in the top-left panel, the in silico sequences sometimes overrepresent deleterious mutations that are easily accessible from the starting nucleotide sequence. 
This is due to the fact that the model does not capture well the effect of such mutations.
Conversely, beneficial mutations that are more distant from the initial nucleotide sequence are underrepresented, as is observed experimentally.

\clearpage

\end{document}